\documentclass{emulateapj}

\shorttitle{An Unveiling Event in NGC 4388}
\shortauthors{M. Elvis et al.}

\begin{document}

\title{An Unveiling Event in the Type 2 AGN NGC4388:\\ 
a Challenge for a Parsec Scale Absorber} 

\author{Martin Elvis\altaffilmark{1}, G. Risaliti\altaffilmark{1,2},
F. Nicastro\altaffilmark{1}, J. M. Miller\altaffilmark{1,3},
F. Fiore\altaffilmark{4}, and S. Puccetti\altaffilmark{4} }

\email{elvis@cfa.harvard.edu}

\altaffiltext{1}{Harvard-Smithsonian CfA, 60 Garden St. 
Cambridge, MA 02138 USA}
\altaffiltext{2}{INAF - Osservatorio di Arcetri, L.go E. Fermi 5,
Firenze, Italy}
\altaffiltext{3}{NSF Astronomy and Astrophysics Postdoctoral Fellow}
\altaffiltext{3}{INAF - Osservatorio di Roma, Sede di Monteporzio
Catone, Via di Frascati, 33, Monteporzio, Italy}
\begin{abstract}
We present two {\em Rossi Xray Timing Explorer} (RXTE) PCA
observations of the type~2 Seyfert galaxy NGC~4388 caught in an
unusual low X-ray absorption state. The observations were triggered by
a detection in the 1.5-3~keV band of the RXTE All-Sky
Monitor. NGC~4388 was found at a somewhat high continuum level
(f(2-10~keV)=8$\times$10$^{-11}$erg~cm$^{-2}$s$^{-1}$) and with a
column density (N$_H\sim$3$\times$10$^{22}$~cm$^{-2}$), a factor
$\sim$10 lower than normal.  The second PCA observation, four hours
later, gave N$_H<$2$\times$10$^{21}$~cm$^{-2}$ indicating, at the
3.1$\sigma$ level, variability so rapid puts the absorber on a few 100
Schwartzschild radii scale, similar to the Broad Emission Line Region,
or smaller. This small scale creates difficulties for the parsec-scale
obscuring torus paradigm of Unified Schemes for type~1 and type~2
AGNs.
\end{abstract}


\section{Introduction}

Optically NGC~4388 is a classical type~2 Seyfert galaxy (Huchra, Wyatt
\& Davis 1982) with permitted and forbidden emission lines of the same
width (Khachikian \& Weedman 1974).  There is abundant evidence that
many, and perhaps all, type~2 Active Galactic Nuclei (AGNs) are normal
type~1 AGNs with both the characteristic broad emission lines and the
optical to X-ray continuum obscured by a flattened torus of absorbing
gas and dust (e.g. Mulchaey et al. 1994). This is the basis of the
Unified Scheme for AGN (Antonucci 1993, Urry \& Padovani 1995).
NGC~4388 has been detected in X-rays for over 20 years (Table~1) and
has always shown a column density, N$_H$= 2-5$\times10^{23}$~cm$^{-2}$.
The most common form of the Unified Scheme locates this absorption in
a dusty torus at parsec distances from the central continuum (Krolik
\& Begelman 1988, Pier \& Krolik 1992, 1993).
However, Risaliti, Elvis \& Nicastro (2002) found that 23/24 X-ray
absorbed AGNs (10$^{22}$~cm$^{-2}<$N$_H<$3$\times$10$^{23}$~cm$^{-2}$)
showed N$_H$ variability by a factor 2-3. The best studied objects
varied on the shortest accessible timescale of months, which is rather
fast to be due to Keplerian motion at parsec radii and so raises
questions about the nature of the obscuring torus. 

Risaliti et al. suggested an alternative location in the cool outer
parts of an accretion disk wind, echoing the model of Kartje,
K{\"o}nigl \& Elitzur (1999) who predicted just such a torus.
This location predicts much faster N$_H$ variability, down to a
timescale of days. In a simple model of Poisson variations in the
number of obscuring clouds, $N_c$, the amplitude of variability found
by Risaliti et al., implies $N_c\sim$5-10. In this case 0.1-1\% of the
time $N_c$=0 and, in the Unified scheme, the central type~1
nucleus would then be unveiled.

The RXTE (Swank et al. 1998) All Sky Monitor (ASM, Remillard et
al. 1997) is just sensitive enough to detect such low energy
`unveiling events'.  We thus began a Target of Opportunity (TOO)
program with Rossi-XTE to obtain snapshot Proportional Counter Array
(PCA, Swank et al. 1998) spectra of type~2 AGNs showing signs of a low
energy detection in the RXTE ASM. Here we report the detection of
a low N$_H$ `unveiling event' in NGC~4388.

\section{Observations and Data Reduction}

We monitored NGC~4388 with the RXTE ASM to search for detections
in the soft 1.5 - 3~keV X-ray channel (`a'). Normally NGC~4388 has a
flux of $\sim$4$\times$10$^{-13}$erg~cm$^{-2}$s$^{-1}$ in this band
(Forman et al. 1979), while a detection requires a flux some 250 times
larger ($\sim$1$\times$10$^{-10}$erg~cm$^{-2}$s$^{-1}$).  Simply
removing the large absorbing N$_H$
($\sim$2-5$\times$10$^{23}$cm$^{-2}$) would increase the observed flux
to $\sim$4$\times$10$^{-11}$erg~cm$^{-2}$s$^{-1}$, a factor of 100, so
that only a modest additional factor 2-3 increase in the emitted
continuum would be needed to put NGC~4388 over the threshold for ASM
detection. By contrast, an increase in the emitted continuum flux by a
factor $>$100 would be unprecedented among the well studied type~1
AGN, where factors of a few to $\sim$10 variation are seen (Markowitz
et al. 2003). Hence an ASM `a' band detection is a good indicator of
a low N$_H$ event.
A triggering event of this type occured on 2003~May~9 (Fig.~1),
shortly after another one (which is visible on the left of Fig.~1).
One day after the ASM trigger NGC~4388 was observed twice with the
RXTE PCA, for 1.9~ksec and 6.2~ksec, with a four hour gap
between the two observations (Fig.~1).

\begin{figure}
\plotone{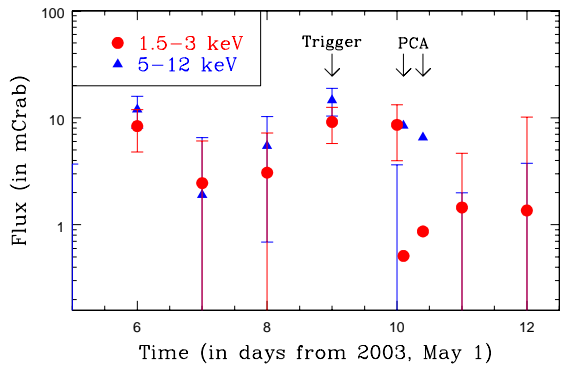} \figcaption{\footnotesize{ The RXTE All-Sky
Monitor (ASM) light curve of NGC 4388 which triggered the pointed
observations. The 'a' (1.5-3~keV) band fluxes are shown as circles and
the 'c' (5-12~keV) band fluxes are shown as triangles . The triggering
data points are marked. The two pointed PCA fluxes in the same bands
are also shown. The error bars on the PCA measurements are smaller
than the points. }}
\end{figure}

We only consider data from PCU-2 for this analysis, as this is the
best-calibrated proportional counter unit (PCU) in the
RXTE/PCA\footnotemark.
\footnotetext{URLs
http://lheawww.gsfc.nasa.gov/users/keith/bkgd\_status/status.html \\
http://heasarc.gsfc.nasa.gov/listserv/grodis/msg00066.html}
Data reduction tools from LHEASOFT version 5.2\footnotemark\ were used
to screen and prepare the event files and spectra.  Data were taken in
"Standard 2" mode, which provides coverage of the full PCA bandpass
(2--60~keV) every 16 seconds.  Only data from the top Xe-filled gas
layer in PCU-2 was used to make the source and background spectra, as
this gas layer has the lowest background.  The standard {\it RXTE}
``good time interval'' filtering was applied to the data.  For the
non-imaging PCA accurate background subtraction is crucial for faint
sources like NGC~4388. The background spectra were made using the tool
{\tt pcabackest} using the latest `faint source' background model
(pca\_bkgd\_cmfaintl7\_e5vv20031123.mdl, updated by {\it RXTE} in
November 2003).  {\tt pcabackest} calculates the predicted (dominant)
particle background every 16~s, and so tracks the variation of the
background around the orbit, thus taking into account the different
backgrounds in these two short observations. Redistribution matrix
files (rmfs) and ancillary response files (arfs) were made and
combined into a single instrumental response file using {\tt pcarsp}.
\footnotetext{URL http://heasarc.gsfc.nasa.gov/docs/software/lheasoft/}
 
We added 0.6\% systematic errors to our spectra using {\tt grppha}, as
we find that in many instances acceptable fits to the Crab can be
obtained with 0.6\% systematic errors.  However, Poisson errors of 5\%
- 10\% are dominant.  The lowest channels in each of the PCUs
routinely reveal strong deviations likely due to calibration
uncertainties; in addition, the calibration of the PCUs is more
uncertain above approximately 25~keV, and the spectra of faint sources
like NGC~4388 become background-dominated in this regime.  In fitting
the spectra, then, we ignored energy range below 3~keV (channels
1-4) and above 20.0~keV.  
%
The PCA X-ray spectra are shown in Fig.~2a, where they are compared
with a {\em Chandra} observation performed 14 months earlier.  Fig.~2a
shows that most of the variation is at low energies, $<$5~keV.  In
Fig.~2b we show the ratio between the two RXTE spectra, which
indicates that the cut off in the spectra at low energies is due to a
difference in N$_H$.

\begin{figure}
\plotone{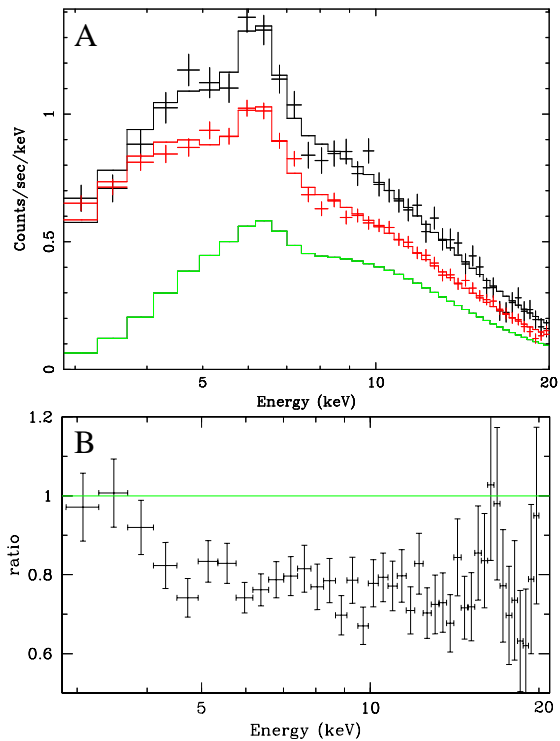} \figcaption{\footnotesize{{\em Upper panel:}
RXTE spectra of NGC 4388. For comparison, we also plot the best
fit model of the {\em Chandra} observation performed one year earlier,
convolved with the response and effective area of the PCA, and
extrapolated from 10~keV to 20~keV for clarity.  {\em Lower panel:}
Ratio PCA-2/PCA-1, showing greater variability at low energies. No
change in N$_H$ would give a flat line.}}
\end{figure}

To fit the PCA spectra we used XSPEC v.11.3\footnotemark. A model
comprising: a power-law of slope $\Gamma$, absorbed by a zero redshift
N$_H$, with a superimposed emission line near the 6.4~keV Fe-K line
energy, was fit to the two PCA data sets and gave good $\chi^2$ (30/34
degrees of freedom for PCA-1 and 25/34 degrees of freedom for
PCA-2. The low redshift of NGC~4388 (2535~km~s$^{-1}$ Huchra, Wyatt \&
Davis 1982) is indistinguishable from zero with the PCA.  The Galactic
N$_H$ (2.5$\times$10$^{20}$cm$^{-2}$, Murphy et al. 1996) is
negligible for the PCA energy range.  The total flux in the kiloparsec
scale X-ray nebula around NGC~4388 (Iwasawa et al. 2003) is also
negligible (2.6$\times$10$^{-13}$erg~cm$^{-2}$s$^{-1}$).  The results
are given in Table~1, together with fits to the same model, with the
same minimum energy, for two unpublished observations from XMM-Newton,
and for six data sets from the literature.  The measured N$_H$ in the
two PCA observations is $\sim5\times10^{22}$~cm$^{-2}$ and
$<0.9\times10^{21}$~cm$^{-2}$, respectively (90\% confidence).  This
rapid change between the two PCA observations in four hours is
significant at the 3.2$\sigma$. I.e. there is a 0.14\% chance that the
N$_H$ value is the same in the two PCA observations.
%
%
\footnotetext{Arnaud K., \& Dorman B., 2003, {\em XSPEC 11.3 User Guide},
URL http://heasarc.gsfc.nasa.gov/docs/xanadu/xspec/manual/manual.html}

The PCA N$_H$ values are one and two orders of magnitude respectively
lower than in all previous X-ray observations of this source. In
particular the PCA-1 N$_H$ is 13$\sigma$ smaller than that measured by
XMM-Newton 5~months earlier (XMM-2, Table~1). [The change in N$_H$ in
the 5 months between XMM-1 and XMM-2 is significant at the 11$\sigma$
level.]
The contour plot of Fig.~3 shows that the dramatic reduction in N$_H$
is not due to a degeneracy between $\Gamma$ and N$_H$ that is
sometimes encountered in X-ray spectra.  $\Gamma$ is flat
($\Gamma$=0.8-0.9) in the PCA spectra compared both with most of the
earlier observations and with unobscured type~1 AGNs ($\Gamma\sim$1.8,
Nandra et al. 1997, Reynolds 1997, Perola et al. 2002). The XMM-1
observation (Table~1) also gave a flat $\Gamma$, yet is heavily
obscured, so a flat spectrum is not a property correlated with a low
N$_H$. The small ($\sim$10~arcsecond) beam size of XMM-Newton (Jansen
et al. 2001) effectively rules out the possibility that the flat PCA
spectral slope is due to another source lying within the
$\sim$1~sq.degree PCA field of view.
The high 2-10~keV flux in the BeppoSAX-1 observation demonstrates that
a similarly high flux does not reduce the observed N$_H$. So we must
be seeing bulk motion across our line of sight, as in Risaliti et
al. (2002).

The N$_H$ variation is robust against reasonable changes in the model:
(1) Adding a reflection component (PEXRAV model, Magdziarz \&
Zdziarski 1995) does not alter $\Gamma$ or $N_H$. Even leaving all the
parameters free, the best fit is obtained with a covering factor
(R=$\Omega/2\pi$) R=0.  The 90\% confidence upper limits are R$<$0.8
for the first observation, and R$<0.2$ for the second one. Fixing R to
these upper limits we obtain column densities $N_{H1}\sim
2\times10^{21}$~cm$^{-2}$ and $N_{H2}\sim8\times10^{22}$~cm$^{-2}$.
(2) Adding a soft component, of course, does change the fit
dramatically as this component affects only the lowest channels. For
black body emission at typical soft excess temperatures of
$kT\sim$0.3-1~keV the required 0.1-2~keV flux is 20~mCrab, or a
luminosity of 5$\times$10$^{42}$erg~s$^{-1}$.  This luminosity causes
no physical problems although a 10$^4$~s variation requires an
optically thick source (Elvis et al. 1991 {\sc ApJ 378, 537}); a
sphere of the implied area has a radius of only 10$^{10}$~cm,
0.3~light-seconds.  However such a large low energy flux is not
suggested by the lowest energy PCA bins that were omitted from the
fit.

\begin{figure}
\plotone{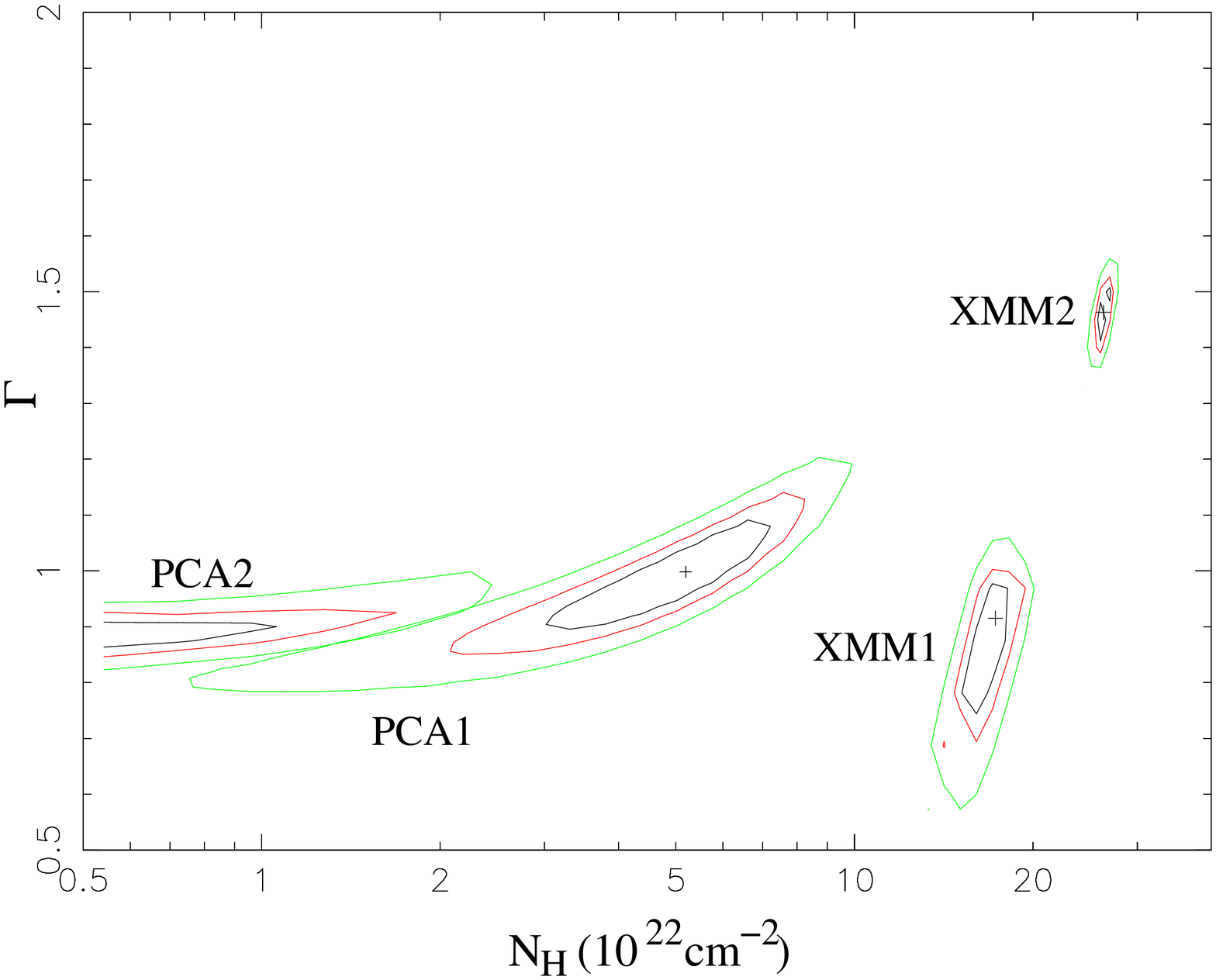} \figcaption{\footnotesize{Photon index ($\Gamma$)
vs. Column density (N$_H$) contour plots (68\%, 90\% and 99\%
confidence), derived here, for the two PCA observations of May 2003,
and the two XMM-Newton observations of July 2002 and December 2002.}}
\end{figure}

The $\chi^2$ test prefers an emission line consistent with the 6.4~keV
Fe-K line, but broadened by $\sigma$=0.42$^{+0.14}_{-0.2}$~keV (90\%
confidence), (i.e. FWHM = 2.38$\sigma$ = 1.0~keV, FWHM/E = 0.15$c$).
A broad line would make an origin for the line in a parsec-scale torus
unlikely, and would argue for wind or accretion disk origin.  The
XMM-2 observation however, using the higher spectral resolution CCD
EPIC detectors, gives a lower gaussian $\sigma$ (73$\pm$20~eV), though
the line may be complex. The PCA-2 measured equivalent width of the
line, $\sim$500~eV, is stronger than normal for a type~1 Seyfert, but
less strong than can be found in Compton thick AGN (Levenson et
al. 2002). Again XMM-2 gives a more normal EW ($\sim$200~eV) for an
AGN that is not Compton thick. The EW and $\sigma$ of the PCA and
XMM-1 are consistent but both disagree with the XMM-2 values.

\begin{table*}
\centerline{\begin{tabular}{lccccccclc} \hline
Instrument & N$_H^a$  &
$\Gamma$ &
E$_{Fe}^b$ &
$\sigma_{Fe}^c$ &
EW$_{Fe}^d$ &
F$^e$ &
L$^f$&
Obs. Date & Ref$^h$ \\
\hline
SL2-XRT  & 2.1$^{+2.8}_{-1.4}$     &  1.9$^{+0.9}_{-0.5}$&--&--&--&
2.1 & 3.5  & 1985 Jul 29 & 1 \\
ASCA-1   & 3.15$^{+1.1}_{-1.0}$    &  1.32$^{+0.77}_{-0.74}$  
&6.49$^{+0.09}_{-0.09}$ & 150$^{+160}_{-130}$ & 750$^{+420}_{-300}$ & 
1.3 & 1.5  & 1994 Jul 04 & 2 \\
ASCA-2   & 3.34$^{+1.0}_{-0.9}$    &  1.47$^{+0.57}_{-0.57}$  
&6.47$^{+0.07}_{-0.07}$ & $<200$ & 700$^{+320}_{-250}$ &
  0.64& 0.7  & 1995 Jun 21 & 2 \\
BeppoSAX-1& 3.80$^{+0.2}_{-0.4}$   &  1.58$^{+0.08}_{-0.22}$  
&6.46$^{+0.07}_{-0.10}$ & $<230$ & 233$^{+115}_{-35}$ &
  2.5 & 3.5  & 1999 Jan 09 & 3 \\
BeppoSAX-2& 4.80$^{+1.8}_{-0.8}$   &  1.47$^{+0.04}_{-0.41}$  
&6.38$^{+0.05}_{-0.06}$ & $<120$ & 525$^{+115}_{-112}$ &
  0.94& 1.4  & 2000 Jan 03 & 3 \\
Chandra  & 3.50$^{+0.4}_{-0.3}$    &  1.8$^j$
&6.36$^{+0.02}_{-0.02}$ & $<230$ & 440$^{+90}_{-90}$ &
  0.36 & 0.6  & 2001 Jun 08 & 4 \\
Chandra-0  & 2.50$^{+0.2}_{-0.1}$    &  1.25$^{+0.14}_{-0.28}$ 
&6.36$^{+0.03}_{-0.03}$ & $<130$ & 165$^{+60}_{-60}$ &
  2.9 & 2.8  & 2002 Mar 05 & 5$^k$\\
XMM-1    & 1.70$^{+0.12}_{-0.15}$  &  0.91$^{+0.10}_{-0.37}$  
&6.41$^{+0.02}_{-0.02}$ & $<77$  & 503$^{+70}_{-60}$ &
  0.77& 0.7  & 2002 Jul 07 & 5 \\
XMM-2    & 2.61$^{+0.06}_{-0.06}$  &  1.46$^{+0.04}_{-0.05}$  
&6.44$^{+0.02}_{-0.02}$ & 73$^{+20}_{-20}$ & 204$^{+24}_{-26}$ &
  2.0 & 2.2  & 2002 Dec 12 & 5 \\
RXTE/PCA-1& 0.52$^{+0.25}_{-0.24}$ &  0.99$^{+0.11}_{-0.11}$  
&6.34$^{+0.11}_{-0.11}$ & $<380$ & 503$_{-100}^{+138}$ &
 7.2 & 4.0  & 2003 May 10$^g$ & 5 \\
RXTE/PCA-2& $<0.09$                &  0.86$^{+0.07}_{-0.03}$  
&6.32$^{+0.10}_{-0.09}$ & 390$^{+160}_{-190}$ & 570$^{+114}_{-100}$ &
  6.1 & 2.9  & 2003 May 10$^g$ & 5 \\
\hline
\end{tabular}}
\caption{\footnotesize{
$^a$ Absorbing column density, N$_H$, in units of 10$^{23}$~cm$^{-2}$.
$^b$ Peak energy of the iron K$\alpha$ line, keV.
$^c$ Width of the iron line, in eV.
$^d$ Equivalent width of the iron line, in eV.
$^e$ Observed 2-10~keV flux, in units of 10$^{-11}$~erg~s$^{-1}$~cm$^{-2}$.
$^f$ Intrinsic 2-10 keV luminosity, in units of
10$^{42}$~erg~s$^{-1}$~cm$^{-2}$ (assuming a Virgo
cluster location at 20 Mpc, Tammann et al. 1999).
$^g$XTE observations 4 hours apart.
$h$ References: 1: Hanson et al. 1990; 2: Forster,
Leighly \& Kay 1999; 3: Risaliti 2002; 4: Iwasawa et al. 2003;
5: This work.
$^j$ Fixed.
$^k$ 0-order spectrum from HETGS observation (Obsid=2983).
}}
\end{table*}

\section{Discussion}

We have found a factor 100 decrease in the column density toward the
normally almost Compton thick ($\tau_{es}\sim$0.1-0.3) type~2 AGN
NGC~4388.  This decrease certainly occured in less than the 0.4~years
from the earlier XMM-Newton observation. The obscuring material in
type~2 AGNs is thus in a highly dynamic state, and warrants intensive
monitoring. A few strongly Compton thick AGN have become almost
Compton thin on a timescale of 2.5 - 5.5 years (Matt, Guainazzi \&
Maiolino 2002), but not on shorter timescales, and still with residual
column densities of $\sim$10$^{23}$cm$^{-2}$ even in the low
absorption state, in 3 out of 4 cases. NGC~4388 is the only known case
of an AGN in which a substantial X-ray opacity changes to an
undetectable value ($\tau_{es}<$0.001).

Moreover, it is likely that the decrease in obscuring column density
coincided with the 2-day `flare' seen in the RXTE ASM, so that
the flare is primarily an `unveiling event'. Without X-ray spectra
through the rise of the ASM flux, however, we cannot be certain.
Between the two PCA observations we saw evidence, at 3.2$\sigma$, for
a change of N$_H$ in four hours.

The probable short timescale (either $\sim$2 days or 4~hours) of the
column density variations has strong implications for the location of
the obscuring matter.  Assuming that the absorption is due to clouds
in Keplerian motion around the central source, and interpreting the
4~hour time lag between the two observations as the crossing time of a
cloud implies a distance from the center R$<10^4 \rho_{10}^2 t_4^2
R_S$, where $\rho_{10}$ is the density in units of
10$^{10}$~cm$^{-3}$, R$_S$ is the Schwartzschild radius, and $t_4$ is
the timescale in units of four hours (Risaliti et al. 2002). If
NGC~4388 has an Eddington ratio of 0.1 then the black hole mass is
$\sim$10$^6$ - 10$^7$M$_{\odot}$. Scaling from the K-band bulge
magnitude gives a similar mass. This implies that the absorber is at
a distance typical of the Broad Emission Line `clouds' (BELRs), or
smaller, and is of similar density.  Only if a high density
($\rho > 10^{12}$~cm$^{-3}$) is assumed can a parsec distant absorber
produce changes on the observed timescale (Fig.~4). A similar
conclusion applies to the Seyfert 1.5 Galaxy NGC~4151, from
the detection of a $\Delta(N_H)\sim2\times10^{23}$ in a time interval
of $\sim150$~ksec during a BeppoSAX observation (Puccetti et al. 2003).

\begin{figure}
\plotone{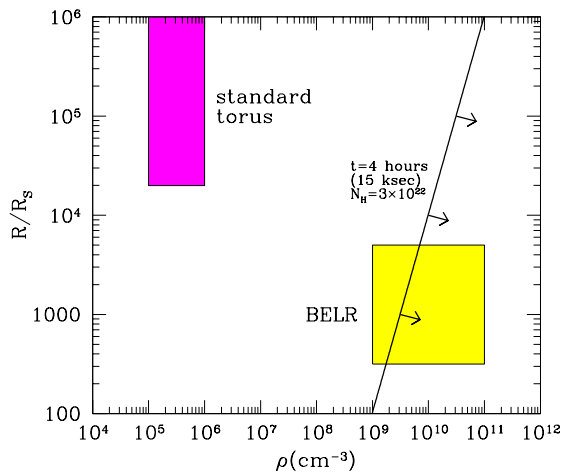} \figcaption{\footnotesize{Distance (R/R$_s$)
vs. density ($\rho$) plot for AGN X-ray absorbers. The
``standard torus'' of the unified model has a density
$n<10^6$~cm$^{-3}$. Variability of the absorber in 4~hours or 
less rules (solid diagonal line) out an absorber with such 
properties, and is more compatible with the standard parameters 
of the Broad Emission Line Clouds (BELR). }}
\end{figure}

This result is a challenge to the parsec scale usually attributed to
the obscuring torus. For spherical isolated clouds, the high density
implies small ($r_c\sim10^{10}$~cm) cloud sizes:
(1) The expansion time for a 100~K ($v_{sound} \sim$ 1~km~s$^{-1}$)
cloud from 10$^{10}$cm to 10$^{11}$cm is only 10$^{6}$~s,, far shorter
than the orbital time. But 
the large pressure needed to confine them
($p/k\sim10^{14}$~K~cm$^{-3}$) cannot be provided by self-gravity.
(2) A hot confining medium would produce thermal emission much greater
than the observed AGN bolometric luminosity (assuming a layer of
1~parsec thickness, L$_{gas}= 10^{48}~T_7~n_7^2$ erg~s$^{-1}$, where
$T_7$ and $n_7$ are the density and temperature in units of 10$^7$~K
and 10$^7$~cm$^{-3}$, respectively, Rybicki \& Lightman 1979).
(3) The N$_H$ through a confining medium would be 10$^{25.5}~n_7\times
d_{pc}$~cm$^{-2}$ =~10$\tau_{es}$. So all continuum variations shorter
than a few years would be smeared out by Thomson scattering, contrary
to observations (Table.1).
(4) Individual clouds would cover only a small fraction,
$\sim$10$^{-6}$, of the X-ray emitting source, whose dimensions are
$>10^{13}$~cm for a black hole of mass $>10^7~M_\odot$ . If hundreds
of clouds are needed to cover the X-ray source, no N$_H$ variation by
more than a few percent can be observed with significant probability.

To produce large variations in N$_H$ compatible with $<N_c>\sim$5-10
needs clouds of a diameter within a factor of a few of the continuum
source size. This implies a density of $\sim$10$^9$cm$^{-3}$. This is
just compatible with a radius $R\sim few$100$R_s$ given a 4~hour
variation (Fig.4). If the absorbers are part of a wind crossing our
line of sight (Elvis 2000) then sheet-like structures (Arav et
al. 1998) become plausible, and this allows larger radii.

The great majority ($>$99\%) of the obscuring {\em gas} in NGC~4388
occurs 
at small radii, and the most likely scenario is that Broad Emission
Line clouds are drifting across our line of sight in NGC~4388 leading
to large changes in N$_H$. Similar behavior has been seen in some
type~1 AGN, involving an N$_H\sim$10$^{23}$cm$^{-2}$ over $\sim$100$^d$
at the radius of the BELR (Lamer, Uttley \& Miller 2003).
Since we have no simultaneous optical spectra we do not know whether
the bulk of the dust, which absorbs the optical and UV photons, lies
on the same small scale. The closest dust could be to the continuum is
the sublimation radius (Barvainis 1987) which, in NGC~4388 at
$\sim$10$^{42}$~erg~s$^{-1}$ (2-10~keV), is $\sim$7$\times$10$^{16}$cm
(0.02~pc), or 2$\times$10$^4 r_s(M_7)$ (fig.4).  The dust:gas ratio in
AGN is typically a factor 10 below the Milky Way value (Maccacaro et
al. 1982, Maiolino et al. 2001), so that a separate dusty absorber is
allowed.
%
%
Unveiling events in type~2 AGN, such as the one reported here for
NGC~4388, put strong constraints on Unified Models for AGN, and seem
to point to a different view of the obscuring torus.


\acknowledgements

This paper used results provided by the ASM/RXTE teams at MIT and at
the RXTE SOF and GOF at NASA's GSFC.  JMM acknowledges the NSF.
This work was partially supported by Chandra grant GO2-3122A.



\end{document}